\documentclass[conference]{IEEEtran}
\IEEEoverridecommandlockouts
\usepackage{cite}
\usepackage{amsmath,amssymb,amsfonts}
\usepackage{algorithmic}
\usepackage{graphicx}
\usepackage{textcomp}
\usepackage{xcolor}
\usepackage{url}
\usepackage{array,multirow}
\usepackage{siunitx}
\usepackage{scalerel,stackengine}
\usepackage{makecell}
\usepackage{multirow}
\usepackage{xcolor}
\usepackage[caption=false, font=footnotesize]{subfig} 
\usepackage[inline]{enumitem}
\usepackage{numprint}

\usepackage{fixme}
\fxsetup{
    status = draft, 
    author = ,
    layout = inline, 
    theme = color}
\definecolor{fxnote}{rgb}{0.0000,0.6000,0.0000} 
\definecolor{fxwarning}{rgb}{1.0000,0.5490,0.0000} 
\definecolor{fxerror}{rgb}{1.0000,0.2706,0.0000}

\graphicspath{{./figures/PNG/}}

\DeclareGraphicsExtensions{.png,.jpg}

\usepackage{glossaries-extra}

\setabbreviationstyle[acronym]{long-short}
\setabbreviationstyle{short-long}
\glssetnoexpandfield{long}
\glssetnoexpandfield{longpl}
\newcommand*{\parentabbr}[2]{%
  \ifglsused{#1}{#2\glsentryshort{#1}}{\protect\glsunset{#1}#2\glsentrylong{#1}}%
}

\newacronym{mtc}{MTC}{Machine Type Communications}
\newacronym{mmtc}{mMTC}{\parentabbr{mtc}{massive }}
\newacronym{m2m}{M2M}{Machine to Machine}
\newacronym{iot}{IoT}{Internet of Things}
\newacronym{nbiot}{NB-IoT}{\parentabbr{iot}{Narrowband }}
\newacronym{gnss}{GNSS}{Global Navigation Satellite Systems}
\newacronym{esa}{ESA}{European Space Agency}
\newacronym{mcs}{MCS}{Modulation and Coding Scheme}
\newacronym{ue}{UE}{User Equipment}
\newacronym{ra}{RA}{Random Access}
\newacronym{rach}{RACH}{Random Access Channel}
\newacronym{nprach}{NPRACH}{Narrowband Physical Random Access Channel}
\newacronym{rv}{RV}{Random Variable}
\newacronym{pdf}{pdf}{probability density function}
\newacronym{gw}{GW}{Gateway}
\newacronym{geo}{GEO}{Geostationary Earth Orbit}
\newacronym{meo}{MEO}{Medium Earth Orbit}
\newacronym{leo}{LEO}{Low Earth Orbit}
\newacronym{fov}{FOV}{Field Of View}
\newacronym{fsl}{FSL}{Free Space Loss}
\newacronym{ul}{UL}{Uplink}
\newacronym{dl}{DL}{Downlink}
\newacronym{db}{dB}{decibel}
\newacronym{snr}{SNR}{Signal-to-Noise Ratio}
\newacronym{sir}{SIR}{Signal-to-Interference Ratio}
\newacronym{sinr}{SINR}{Signal-to-Interference plus Noise}
\newacronym{ecef}{ECEF}{Earth Centered Eart Fixed}
\newacronym{kpi}{KPI}{Key Performance Indicator}
\newacronym{phy}{PHY}{physical layer}
\newacronym{lut}{LUT}{Look Up Table}
\newacronym{awgn}{AWGN}{Additive White Gaussian Noise}
\newacronym{3gpp}{3GPP}{Third Generation Partnership Project}
\newacronym{nr}{NR}{New Radio}
\newacronym{ntn}{NTN}{Non Terrestrial Network}
\newacronym{ml}{ML}{Machine Learning}
\newacronym{uas}{UAS}{Unmanned Aircraft System}
\newacronym{isl}{ISL}{Inter-Satellite Links}
\newacronym{nn}{NN}{Neural Networks}

\def\BibTeX{{\rm B\kern-.05em{\sc i\kern-.025em b}\kern-.08em
    T\kern-.1667em\lower.7ex\hbox{E}\kern-.125emX}}
\begin{document}

\title{Location-assisted precoding in 5G LEO systems: architectures and performances}

\author{\IEEEauthorblockN{Alessandro Guidotti\IEEEauthorrefmark{1}, Carla Amatetti\IEEEauthorrefmark{2}, Fabrice Arnal \IEEEauthorrefmark{3}, Baptiste Chamaillard\IEEEauthorrefmark{3}, and
Alessandro Vanelli-Coralli\IEEEauthorrefmark{2}}
\IEEEauthorrefmark{1}National Inter-University Consortium for Telecommunications (CNIT), Bologna, Italy
\\
\IEEEauthorblockA{\IEEEauthorrefmark{2}Dept. of Electrical, Electronic, and Information Engineering (DEI), Univ. of Bologna, Bologna, Italy}
\IEEEauthorblockA{\IEEEauthorrefmark{3}Thales Alenia Space, 26 Avenue JF Champollion, 31000 Toulouse, France}}
\maketitle

\begin{abstract}
Satellite  communication  systems  are  a fundamental component in support of Europe’s ambition to deploy smart and sustainable networks and services for the success  of  its  digital  economy.  To cope with the 5G and beyond ever increasing demand for larger throughput, aggressive frequency reuse schemes (\emph{i.e.}, full frequency reuse), with the implementation of precoding/beamforming to cope with the massive co-channel interference, are recognised as one of the key technologies. While the best performance can be obtained with the knowledge of the Channel State Information (CSI) at the transmitter, this also poses some technical challenges related to signalling and synchronisation. In this paper, we focus on precoding solutions that only needs the knowledge of the users' positions at the transmitter side, namely the recently introduced Switchable Multi-Beam (MB) and Spatially Sampled MMSE (SS-MMSE) precoding. Compared to the vast majority of the studies in the literature, we take into account both the users' and the satellite movement in a Low Earth Orbit (LEO) mega-constellation, also proposing two system architectures. The extensive numerical assessment provides a valuable insight on the performance of these two precoding schemes compared to the optimal MMSE solution.  
\end{abstract}

\glsresetall

\section{Introduction}\label{sec:intro}
Satellite systems are expected to play a crucial role in future wireless networks. The inclusion of the Non-Terrestrial Network (NTN) in 3GPP Rel. 17 will improve the system flexibility, adaptability, and resilience, and extend the 5G coverage to rural and under/un-served areas. To completely enable this new role of Satellite Communication (SatCom) systems, it is necessary to satisfy the user demand, which, in the last few years, has become more and more heterogeneous in terms of  services (\emph{e.g.}, Internet of Things (IoT), Mission Critical communications, and enhanced mobile broadband) characterised by very different performance requirements concerning rate and delays. In order to meet the 5G requirements, both academia and industry have been focusing on advanced system-level techniques to increase the offered capacity. One possible way to reach it is the exploitation of the available spectrum bandwidth, by either adding unused or underused spectrum chunks by means of flexible spectrum usage paradigms (\emph{e.g.}, Cognitive Radio solutions, \cite{icolari2015interference,chatzinotas2017cognitive,liolis2013cognitive}) or by fully exploiting the spectrum by decreasing the frequency reuse factor down to full frequency reuse (FFR). With the latter, high co-channel interference from adjacent beams is introduced, which requires the adoption of sophisticated interference management techniques, either at transmitter-side, \emph{e.g.}, precoding \cite{arapoglou2010mimo,YoLiWa20,guidotti2020clustering,guidotti2021design,8510728,mosquera_2021,gallinaro2005perspectives}, or at receiver-side, \emph{e.g.}, Multi-User Detection (MUD) \cite{colavolpe2016multiuser}. During the last years, the implementation of beamforming techniques in SatCom has been extensively addressed for Geostationary Earth Orbit (GEO) systems, mainly, but also for Low Earth Orbit (LEO) constellations, as reported in \cite{arapoglou2010mimo,YoLiWa20,guidotti2020clustering,guidotti2021design,8510728,mosquera_2021,gallinaro2005perspectives} and the references therein. In these works, the objective has been that of increasing the overall throughput in unicast or multicast systems, also addressing well-known issues for SatCom-based beamforming as scheduling and Channel State Information (CSI) retrieval. Finally, the design of hybrid beamforming for massive Multiple Input Multiple Output (MIMO) communications in LEO systems has been recently addressed in, \cite{mosquera_2021}; here, the authors focus on a specific implementation of an on-board beamforming codebook compatible with 3GPP New Radio (NR). A thorough survey on MIMO techniques applied to SatCom is provided in \cite{arapoglou2010mimo}, where both fixed and mobile satellite systems are examined and the major impairments related to the channel are identified. Notably, a critical challenge is the availability of CSI at the transmitter (CSIT), especially in systems involving Non Geostationary Satellites (NGSO). Such problem is also exacerbated by the mobility of both the UEs and the satellites, which can make the coherence time of the channel shorter than the transmission delay. The impact of non-ideal CSI at the transmitter, when applying precoding to a SatCom context are discussed in \cite{zorba2008improved}, where, the authors propose a novel MIMO scheme aimed at increasing the system sum-rate, availability, and variance performance.
\\
In order to avoid/limit the need for the CSI reporting to the transmitter, in this paper we focus on precoding techniques which only require the knowledge of the users’ positions,\emph{i.e.}, Multi-Beam  (MB) precoding, \cite{angeletti2020pragmatic}, and propose a novel algorithm, based on the Minimum Mean Squared Error (MMSE) approach, which does however not need CSIT,  denoted as Spatially Sampled MMSE (SS-MMSE) precoding. Two system architectures are discussed, differentiated by where the precoding coefficients are computed based on the selected functional split option. Moreover, differently from many other works, both the UEs and the satellite movement are considered. 
\\
The remainder of the work is the following: in Section II the system architecture is described, Section III outlines the system model and the assumptions, in  Section IV we provide the numerical assessment and a detailed discussion about the results. Finally, Section V concludes this work.

\begin{figure}[t!]
    \centering
    \includegraphics[width=\columnwidth]{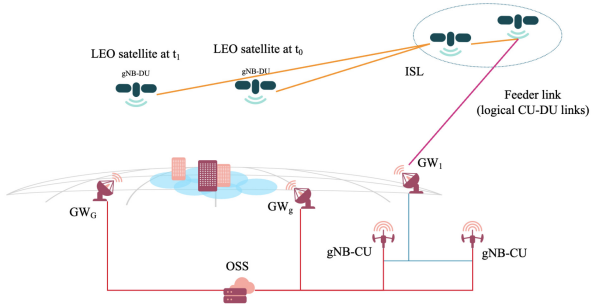}
    \caption{System architecture for 5G precoding with a single LEO satellite.}
    \label{fig:architecture}
\end{figure}

\section{System architecture}
There are several design choices that impact the definition of the architecture when precoding and beamforming are considered; among them, we focus on: i) the type of NR gNodeBs (gNB)  functional  split  that  is  implemented,  if  any,  as  per  3GPP TR 38.801,  \cite{3gpp_38801_2017};  ii) the network entity in which the precoding coefficients are computed; and iii) the network entity in which the coefficients are applied to the signals.
Referring to Fig.~\ref{fig:architecture}, the system architecture is composed by:
\begin{itemize}
    \item The terrestrial segment, where the terrestrial data networks are connected to the NTN segment through a set of on-ground Gateways (GWs). The latter provide inter-connectivity between  the  satellite  constellation, the gNBs, and the Core Network (CN) through the ground distribution network, in particular with the Operations Support Systems (OSS) entity, in charge of managing the overall system.
    \item The access segment is assumed to be provided by regenerative LEO satellites, whose coverage can be achieved with \textit{fixed} or \textit{moving} beams.  In the former case, the on-board antenna keeps serving the same on-ground area while the satellite moves on its orbit (steerable antennas). In the latter case, the served on-ground area is moving together with the satellite. Inter-Satellite Links (ISLs) are exploited to provide a logical link between the LEO satellite and the serving gNB on-ground, since they might not always be in direct visibility.
    \item The on-ground user segment, composed by a potentially massive number of users distributed all over the world. The UEs are assumed to be directly connected to the NGSO node by means of the Uu air-interface through the user access link. 
\end{itemize}
With functional split, the gNB can be split in: 1) a Central Unit (gNB-CU), \emph{i.e.}, a logical node that provides support for the upper layers of the protocol stack (\emph{e.g.}, for mobility control, radio access network sharing, positioning, session management, etc.); and 2) a Distributed Unit (gNB-DU), \emph{i.e.}, a logical node that includes  the gNB lower layers,  such  as  Layer 1 and 2 functions. It shall be noticed that a single on-ground gNB-CU can manage multiple on-board gNB-DUs. In general, for the purpose of this work related to the implementation of precoding techniques, the main difference in the functional split options  is related  to  where,  between  the  gNB-DU and the gNB-CU, the scheduling and the precoding coefficients are computed. Based on this design choice, we categorise the architecture as follows: i) \textit{Centralised  Precoding  Computation  (CPC)}, where  scheduling and precoding are computed at the on-ground gNB-CU; and ii) \textit{Distributed  Precoding  Computation  (DPC)}, where the  functional  split  is selected to implement on-board the computation of the scheduling and precoding matrices. With CSI-based algorithms, the choice between CPC and DPC is critical. With the latter, the CSI vectors estimated by the users are provided to the satellite, which computes the precoding matrix and transmits the data; with the former, the CSI vectors shall be sent back to the on-ground gNB and then the precoding coefficients shall be sent to the satellite, increasing the time interval between when the CSIs are computed (estimation phase) and when the corresponding precoding matrix is used to transmit the data (transmission phase). However, it shall also be mentioned that DPC requires more complex payloads, since more layers must be implemented on-board.

For the MB and SS-MMSE solutions, introduced below, the CSIs are not needed; however, the users shall provide their location, obtained by means of Global Navigation Satellite System (GNSS) capabilities, which can be assumed for the majority of NTN UEs. Finally, all algorithms (CSI and non-CSI based) require the knowledge of the UEs' capacity request  and  type  of  traffic,  so  as  to  fed  them  to  the  Radio Resource Management (RRM)  algorithm, and the terminal type, so as to include the noise power levels in the precoding equations, \emph{e.g.}, handheld or Very Small Aperture Terminal (VSAT). With respect to the latter, it shall be mentioned that this information might be classified by the manufacturers; in this case, an estimate can be identified based on ancillary terminal parameters/information. 

\section{System model}
In the following, we focus on a single LEO satellite with moving beams providing connectivity to $N_{UT}$ uniformly distributed on-ground UEs by means of $N_{B}$ beams generated by an on-board planar antenna array with $N_{F}$ radiating elements. As  previously introduced,  the  considered  precoding  algorithms  require  either  the  CSI provided  by  the  UEs (MMSE)  or  their  location (MB, SS-MMSE) in  order  to  compute  the precoding  matrix. These  values  are  computed  by  the  users at a time instant $t_0$ (see Fig.~\ref{fig:architecture}); the precoding matrix is then computed at the gNB-CU (CPC) or by the gNB-DU (DPC) and, then, actually used to transmit  the  precoded  symbols  to  the  users  at  a  time $t_1$. The  latency $\Delta t = t_1 - t_0$ between   the   estimation   and   the   transmission   phases     introduces   a misalignment  in  the  channel  to which  the precoding  matrix  is  matched and  the  channel that   is   actually   encountered   during   the   transmission,   thus   impacting   the   system performance. Thus, the delay between the estimation instant and that in which precoding actually happens is given by:
\begin{equation}
    \Delta t = t_{ut,max} + 2t_{feeder} + t_p + t_{ad}
\end{equation}
where $t_{ut,max}$ is  the  maximum propagation delay for  the  user  terminals  requesting connectivity  in  the  coverage  area, $t_{feeder}$ is  the  delay  on  the  feeder  link between  the satellite  connected  to  the  GW  (and,  thus,  to  the  reference gNB-CU for CPC), $t_p$ is the processing delay to compute the precoding matrix, and $t_{ad}$ includes additional delays, as that between the estimation and its reporting. When DPC is implemented, the latency to obtain the users’ information and compute the precoding matrix is given by $t_{ut,max}+ t_p$ only; however, in order to also obtain the users’ symbols to be precoded, the other terms have to be considered and, thus, no significant difference arises between CPC and DPC from this point of view. It shall be noticed that, in this time period,  there  are  several  sources  of  misalignment  between  the  channel  coefficients  or locations estimated to compute the precoding matrix and the channel realisation when the precoded transmission occurs: i) the satellite moved along its orbit; ii) the user terminals might  have  moved  depending  on  the  terminal  type;  iii)  different realisations  of  the stochastic  terms  representing the  additional  losses  (\emph{e.g.},  large  scale  loss, scintillation)  are  present. Assuming  FFR,  the CSI vector  at  feed  level, $\mathbf{h}_i^{(feed)} = [h_{i,1}^{(feed)},\ldots,h_{i,N_{F}}^{(feed)}]$ represents  the channel between the $N_F$ radiating elements and the generic $i$-th on-ground user terminal, $i=1,\ldots, N_{UT}$:
\begin{equation}
    h_{i,n}^{(feed)} = \frac{g_{i,n}^{(tx)} g_{i,n}^{(rx)}}{4\pi\frac{d_{i}}{\lambda}\sqrt{L_i\kappa BT_i}} e^{-\jmath \frac{2\pi}{\lambda} d_{i}} , \ n=1,\ldots,N_F
\end{equation}
where: i)  $d_{i}$ is the slant range between the $i$-th user and the antenna feeds, which for a single satellite can be assumed to be co-located; ii) $\lambda$ is the wavelength; iii) $\kappa BT_i$ denotes the equivalent thermal noise power, with $\kappa$ being the Boltzmann constant, $B$ the user bandwidth (for simplicity assumed to be the same for all users), and $T_i$ the equivalent noise temperature of the $i-th$ user receiving equipment; iv) $L_{i}$ denotes the additional losses considered  between  the $i$-th  user  and  the co-located  antenna  feeds; and v) $g_{i,n}^{(tx)}$ and $g_{i,n}^{(rx)}$ denote the transmitting and receiving complex antenna patterns between the $i$-th user and the $n$-th antenna feed. The additional losses are computed as $L_{i} = L_{sha,i} + L_{atm,i} + L_{sci,i} + L_{CL,i}$, where $L_{sha,i}$ represents the log-normal shadow fading term, $L_{atm,i}$ the atmospheric loss, $L_{sci,i}$ the scintillation, and $L_{CL,i}$ the Clutter Loss (CL); these terms are computed as per 3GPP TR 38.821. Collecting  all  of  the $N_{UT}$ CSI vectors, it is possible to build a $N_{UT} \times N_F$ complex channel matrix at system level $\mathbf{H}_{sys}^{(feed)},$ where the generic $i$-th row contains the CSI vector of the $i$-th user and the generic $n$-th column contains the channel coefficients from the $n$-th on-board feed towards  the $N_{UT}$ on-ground  users. During  each  time  frame,  the  RRM algorithm (which is out of the scope of this work) identifies a subset of $N_{sch}$ users to be served, leading to a $N_{sch} \times N_F$ complex scheduled channel matrix $\mathbf{H}^{(feed)} = \mathcal{S} \left(\mathbf{H}_{sys}^{(feed)}\right)$, where $\mathcal{S}$ denotes the RRM scheduling function, which is a sub-matrix of $\mathbf{H}_{sys}^{(feed)}$, \emph{i.e.}, $\mathbf{H}^{(feed)} \subseteq \mathbf{H}_{sys}^{(feed)}$, which contains only the rows of the scheduled users.
The  selected  precoding algorithm computes a $N_{sch}  \times N_F$ complex precoding matrix $\mathbf{W}$ which projects the $N_{sch}$ dimensional column  vector $\mathbf{s} = [s_1,..,s_{N_{sch}}]^T$ containing the unit-variance user  symbols onto  the $N_F$-dimensional  space defined by the antenna feeds. Thus, in the feed space, the beamforming and precoding matrices are jointly computed, allowing for the generation of a dedicated beam towards each user direction. The signal received by the $k$-th user can be expressed as follows: 
\begin{equation}
    y_k = \underbrace{\mathbf{h}_{k,:}^{(feed)}\mathbf{w}_{:,k}s_k}_\text{intended}  + \underbrace{\sum_{\substack{i=1\\ i\neq k}}^{N_{sch}}\mathbf{h}_{i,:}^{(feed)}\mathbf{w}_{:,i}s_i}_\text{interfering} + z_k
\end{equation}
where $z_k$ is a circularly symmetric Gaussian random variable with zero mean and unit variance, this is legit observing that the channel coefficients in (2) are normalised to the noise power.
The $N_{sch}$-dimensional vector of received symbols is:
\begin{equation}
    \mathbf{y} = \mathbf{H}_{t_{1}}^{(feed)}\mathbf{W}_{t_{0}}\mathbf{s}+\mathbf{z}
\end{equation}
Note  that,  as  previously  discussed,  the  channel  matrix,  that  is  used to  compute  the precoding matrix, is referring to a time instant $t_0$, while the precoded symbols are sent to the users at a time instant $t_1$, in which the channel matrix will be different.
When  considering the beam  space  precoding, the  beamforming  and precoding  matrices are  distinct,  although  they  can still  be jointly  optimised.  In particular,  first a desired beam lattice on-ground is defined in order to generate $N_B$ beams, with $\mathbf{c}_{\ell}$ denoting the $(u,v)$ coordinates  of  the  generic ${\ell}$-th  beam, ${\ell}=1,\ldots,N_B$. The $N_F \times N_B $ complex  beamforming matrix $\mathbf{B}$ generates an equivalent channel in the beam space by linearly combining the signals emitted by the $N_F$ antenna feeds, \emph{i.e.}, $ \mathbf{H}^{(beam)} = \mathbf{H}^{(feed)}\mathbf{B}$, where the $k$-th row of the beam channel matrix $\mathbf{H}^{(beam)}$, $\mathbf{h}_{k,:}^{(beam)}$, provides the equivalent channel coefficients of the $k$-th on-ground user. The $N_F$-dimensional beamforming column vector steering the radiation pattern towards the ${\ell}$-th beam center can be computed as
\begin{equation}
    \mathbf{b}_{:,{\ell}} = [b_{1,{\ell}},..,b_{N_F,{\ell}}], \; \mathrm{with} \; b_{n,{\ell}} = \frac{1}{\sqrt{N_F}}e^{-jk_0\mathbf{r}_n\cdot \mathbf{c}_l}
\end{equation}
where $\mathbf{r}_n$ is the position of the $n-th$ array element with  respect  to  the  antenna  center. Exploiting (4), the received signal is given by
\begin{equation}
    \mathbf{y} = \mathbf{H}_{t_{1}}^{(beam)}\mathbf{W}_{t_{0}}\mathbf{s}+\mathbf{z} = \mathbf{H}_{t_{1}}^{(feed)}\mathbf{B}\mathbf{W}_{t_{0}}\mathbf{s}+\mathbf{z}
\end{equation}
In terms of precoding schemes, the MB algorithm is based on a pre-computed codebook, \cite{AnDe20}, in which each user is associated to the closest beam center and precoded with the corresponding beamforming vector. Thus, assuming that one user from each beam is served at each time-slot,  $\mathbf{W}_{MB}=\mathbf{B}$. This approach is simple and computationally effective; however, a better performance can be achieved by observing that, for a given user location, additional information can be obtained. In the proposed SS-MMSE algorithm, the CSI vectors are not estimated by the users but approximated at the transmitter side in the directions of the beam centers (BC):
\begin{equation}
    \widehat{\mathbf{h}}_{i,n}^{(feed)} = \frac{g_{i,n}^{(tx,BC)} g_{i,n}^{(rx,BC)}}{4\pi\frac{d_{i}^{(BS)}}{\lambda}\sqrt{\kappa BT_i}} e^{-\jmath \frac{2\pi}{\lambda} d_{i}^{(BS)}}, \ n=1,\ldots,N_F
\end{equation}
which is obtained from (2) by excluding all terms that are not known based on the beam center location, \emph{i.e},  the additional losses.  The terms in the approximated channel coefficient can be obtained based on the user location and the satellite ephemeris. The CSI vectors obtained with this approach can then be fed to the well known MMSE precoding algorithm:
\begin{equation}
    \mathbf{W}_{SS-MMSE} = \widehat{\mathbf{H}}^H(\widehat{\mathbf{H}}^H \widehat{\mathbf{H}} + \mathrm{diag}(\boldsymbol{\alpha})I_{N_B})^{-1} \widehat{\mathbf{H}}^H
\end{equation}
where $\widehat{\mathbf{H}}$ is the estimated channel matrix in the beam or feed space. In the above equation, $\boldsymbol{\alpha}$ is a  vector of  regularisation factors, with optimal value given by the inverse of the expected Signal-to-Noise Ratio (SNR) on the link. Finally, as extensively detailed in \cite{guidotti2021design}, the power normalisation is a fundamental step for precoding and beamforming so as to properly take into account the power that can be emitted both by the satellite and per antenna: i) with the Sum Power Constraint (SPC), an upper bound is imposed on the total on-board power as $\widetilde{\mathbf{{W}}} = \frac{\sqrt{P_{t}}\mathbf{{W}}}{\sqrt{\mathrm{tr}\left(\mathbf{W}\mathbf{W}^{\mathit{H}}\right)}}$, with $P_t$ being the total on-board power, which preserves the orthogonality of the precoder columns but does not guarantee that the power transmitted from each feed will be upper bounded, \emph{i.e.}, it might be working in non-linear regime; ii) with Per Antenna Constraint (PAC), the limitation is imposed per antenna with $\widetilde{\mathbf{{W}}} = \sqrt{\frac{P_{t}}{{N}}}\mathrm{diag}\left(\frac{1}{\parallel {{\mathbf{w}}}_{{1},:} \parallel },\ldots,\frac{1}{\parallel {\mathbf{w}}_{\mathit{N},:}\parallel}\right)\mathbf{\mathrm{W}}$ ($N=N_F,N_B$ for feed or beem space precoding), but the orthogonality in the precoder columns is disrupted; and iii) with the Maximum Power Constraint (MPC) solution, $\widetilde{\mathbf{{W}}} = \frac{\sqrt{P_{t}}\mathbf{{W}}}{\sqrt{{N\max_{j}{\parallel\mathbf{{w}}_{\mathit{j},:} \parallel}^2}}}$, the power per antenna is upper bounded and the orthogonality is preserved, but not the entire available on-board power is exploited. In this framework, it is straightforward to notice that with the MB algorithms the three normalisations lead to the same precoding matrix, since the beamforming vectors are normalised by definition in (5).
	
\section{Numerical results and discussion}
In this section, we report the outcomes of the extensive numerical assessment configured as reported in Table~\ref{tab:parameters}, considering a single LEO satellite at $600$ km. Both fixed and public safety terminals are considered and the following Key Performance Indicators (KPIs) are computed and evaluated: average values and Cumulative Distribution Functions (CDFs) of Signal to Interference plus Noise Ratio (SINR), Signal to Interference Ratio (SIR) and achievable spectral efficiency. While the user density might seem limited, it shall be recalled that we are not considering scheduling algorithms and, thus, the user density does not impact the overall performance, as long as the number of Monte Carlo iterations guarantees the system convergence. The assessment is performed in full buffer conditions, \emph{i.e.}, infinite traffic demand. Based on these assumption, the users are randomly scheduled. In particular, at each time frame one user from each beam is randomly selected to be served and the total number of time frames is computed so as to guarantee that all users are served. The numerical assessment is provided with MB and SS-MMSE precoding and the performance benchmark is the one obtained with MMSE precoding and ideal CSI estimates at the transmitter side.  
\subsubsection{Fixed terminals}
\begin{figure}[t!]
    \centering
    \subfloat[Feed space \label{fig:plosbeam}]{%
        \includegraphics[width=0.5\linewidth]{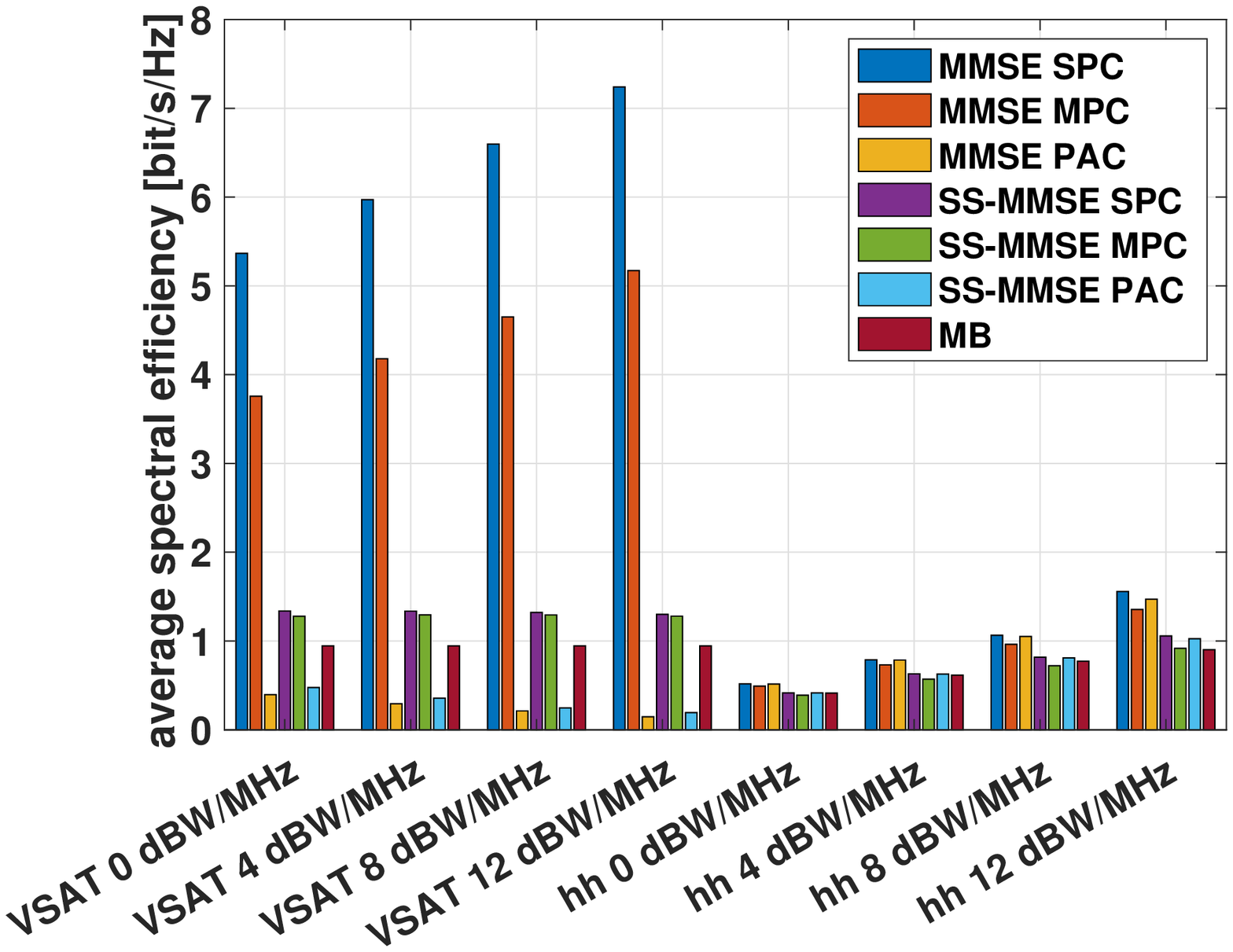}}
    \hfill
    \subfloat[Beam space\label{fig:plosfeed}]{%
        \includegraphics[width=0.5\linewidth]{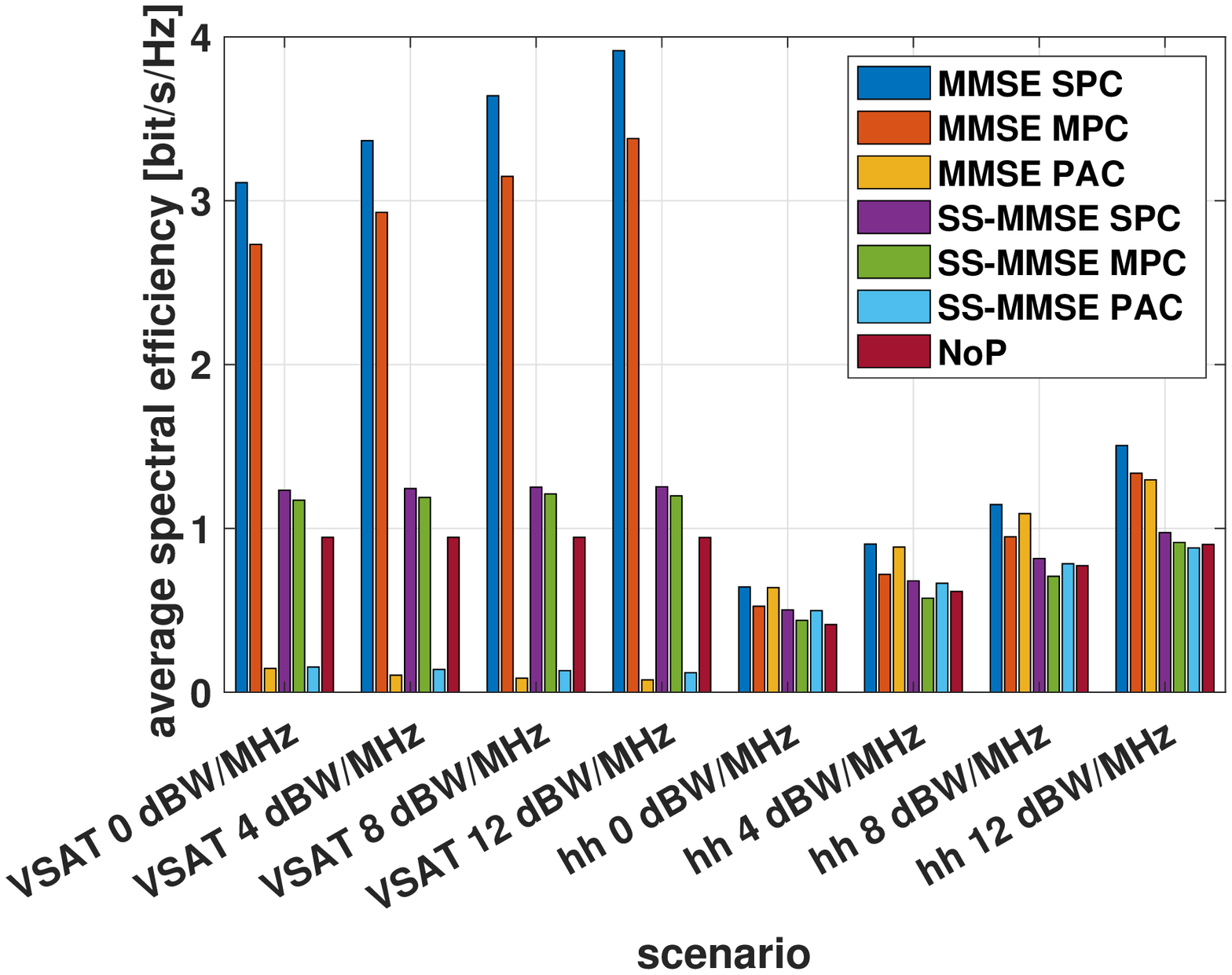}}
  \caption{Average Spectral efficiency of fixed users in pLoS scenario.}
  \label{fig:analysis} 
\end{figure}

\begin{figure}[t!]
    \centering
    \subfloat[SINR \label{fig:sinr}]{%
        \includegraphics[width=0.5\linewidth]{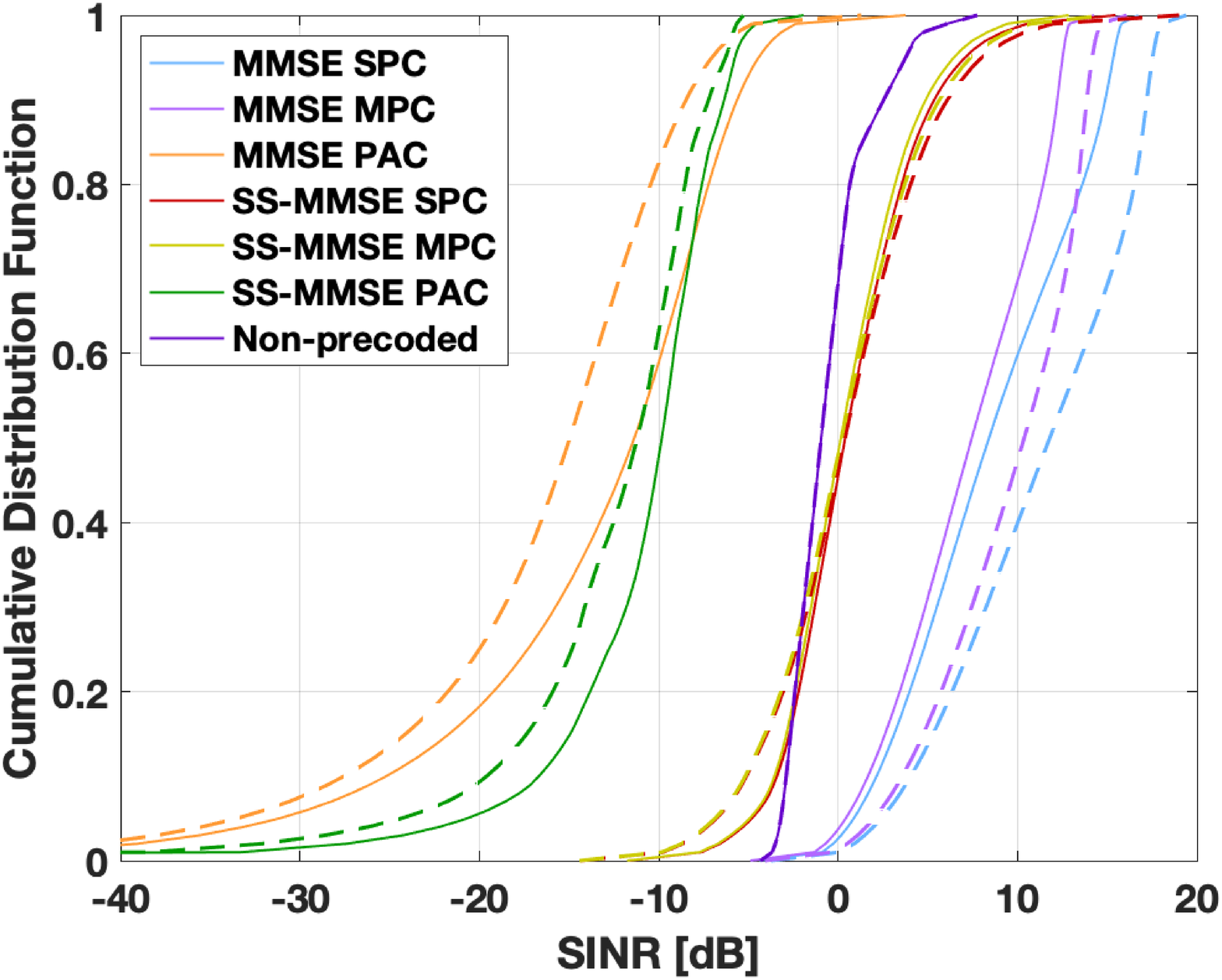}}
    \hfill
    \subfloat[SIR\label{fig:sir}]{%
        \includegraphics[width=0.5\linewidth]{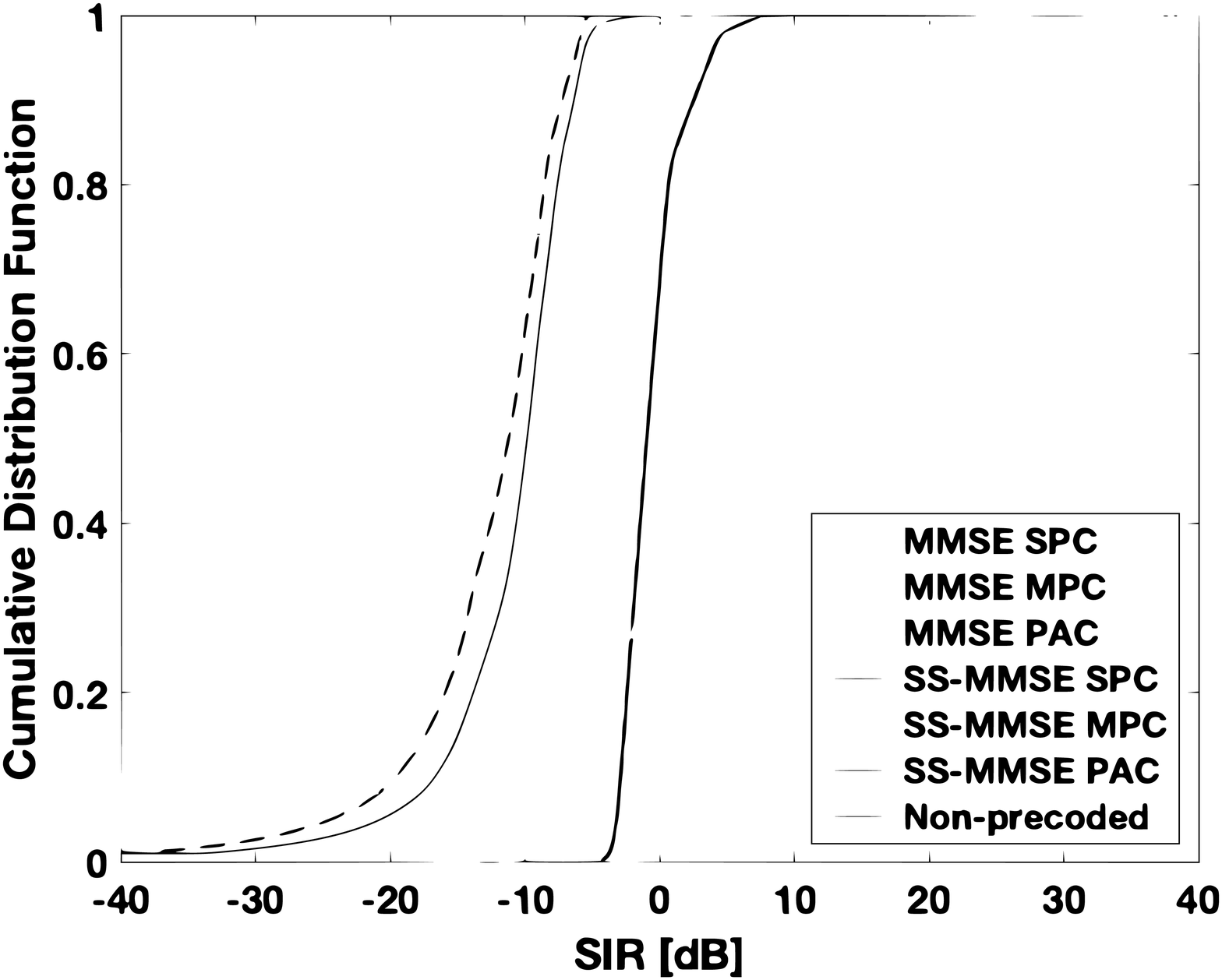}}
  \caption{SINR, and SIR CDFs for VSAT terminals in the pLOS scenario for beam space precoding, $P_t = 0 dBW/MHz$ (solid line) and $P_t = 12 dBW/MHz$ (dashed line).}
  \label{fig:analysis} 
\end{figure}

we first focus on the pure LOS (pLOS) scenario, in which the channel coefficients do not include any additional loss as per TR 38.821 \cite{3gpp_38821_2021} and TR 38.811 \cite{3gpp_38811_2020}, but it only accounts for free space loss, noise, and phase rotation due to the slant range. Figure ~\ref{fig:plosbeam} reports the corresponding histograms of the average spectral efficiency when precoding in the beam space is applied. In general, it is possible to observe that the MMSE precoding provides a better performance compared to SS-MMSE and the non-precoded scenario, as expected. However, with low transmitted power and handheld terminals the SS-MMSE approach is relatively close to the performance of MMSE. This is motivated by observing that, when the power increases and in particular with VSAT terminals that have a large receiving antenna gain, there is a more critical need for a better interference limitation to avoid any approximation in the precoding matrix, and thus the MMSE precoder provides significantly better results. In scenarios with a reduced need for interference limitation, the SS-MMSE is a good solution. In terms of normalisations, SPC always provides the best performance as expected. However, this approach does not guarantee that an antenna or feed does not exceed the power it can emit and, thus, the MPC and PAC solutions should be preferred. Comparing them, it can be noticed that the MPC is significantly better when the interference in the system is larger, \emph{i.e.}, for large transmission power and VSAT terminals with large antenna gains: in this case, it is fundamental to keep the orthogonality in the precoding matrix columns. With handheld terminals, both for MMSE and SS-MMSE, as long as the power is limited, it is more important to increase the SNR and, thus, PAC is better. This solution guarantees that each feed or antenna emits the same power level, while perturbing the precoding orthogonality. When the power is increased, interference becomes more impacting and MPC is again the best option. Comparing the two considered user equipment types, VSATs provide a much better performance thanks to the significantly larger antenna gain compared to handheld terminals. In this scenario, it is worth noticing that there is no advantage of VSATs related to interference rejection with the directive radiation pattern, since it is assumed that all of the UEs’ antennas are pointed towards the single satellite, with the legit assumption of co-located antenna feeds. Finally, observing the trends as a function of the transmission power, a larger power allocation leads to larger average rate values. However, this does not apply for VSAT terminals in the absence of precoding, indeed, in this case, the intended and interfering power levels change accordingly and, as a consequence, the SINR level is almost constant, with a slight decrease at $P_t=12 dBW/MHz$. With handheld terminals, more limited in terms of receiving antenna gain, larger power levels lead to larger spectral efficiencies.
The above trends are substantiated by the results shown in Figures ~\ref{fig:sinr} and ~\ref{fig:sir}, which reports the CDFs for the and SINR and SIR in the pLOS scenario for VSAT terminals in the beam space. It can be noticed that with SPC and for increasing transmission power levels, the SIR increases accordingly, leading to a better SINR. As for PAC, a larger transmission power leads to a worse SINR curve, denoting a significant sensitivity to the loss of orthogonality in the precoding matrix columns in scenarios with increased interference. Looking at figure ~\ref{fig:sir}, MPC and SPC have a significantly better performance in limiting interference compared to both the non-precoded and PAC cases. Actually, the PAC normalisation leads to a performance that is even worse than the non-precoded case with VSATs, highlighting the poor interference rejection obtained with this approach in scenarios with a significant co-channel interference. It is also worth mentioning that, for MPC and SPC, the SIR plots are overlapped. Indeed, the SIR does not depend on a scalar multiplicative factor and, consequently, it is exactly the same in both normalisations.

\begin{table}[t!]
\tiny
\centering
\caption{Simulation parameters}
\label{tab:parameters}
 \begin{tabular}{|c|c|} 
 \hline
 \thead{Parameter} & \thead{Range}\\
 \hline\hline
 System band & S ($2$ GHz) \\ 
\hline
Beamforming space & feed,beam \\
\hline
Receiver type & VSAT, handheld (hh) \\
\hline
Receiver scenario & fixed, public safety \\
\hline
Propagation scenario & pLOS, NLOS \\
\hline
Total on-board power density, $P_{t,dens}$ & $0,4,8,12$ dBW/MHz \\
\hline
Number of beams $N_b$ & $91$ \\
\hline
User density & $0.5$ $user/km^2$ \\
\hline
Monte Carlo iterations & $70$ \\
\hline
 \end{tabular}
\end{table}

Figure ~\ref{fig:plosfeed} reports the results for feed space precoding, in which MB precoding is included. As for the beam space, the MMSE precoding is always providing the best performance, followed by the SS-MMSE approach. However, while this is always true for the SPC and MPC normalisations, when PAC is considered the MB precoding is better due to the loss in terms of interference limitation of the PAC normalisation which leads to a better performance implementing beamforming only (MB). The performance of precoding in the feed space is better for larger power levels as long as the SPC and MPC normalisations are used with VSATs and in all cases for handheld terminals. However, when PAC is used for VSATs, the performance becomes worse. 

\begin{figure}[t!]
    \centering
    \subfloat[Feed space \label{fig:nlosfeed}]{%
        \includegraphics[width=0.5\linewidth]{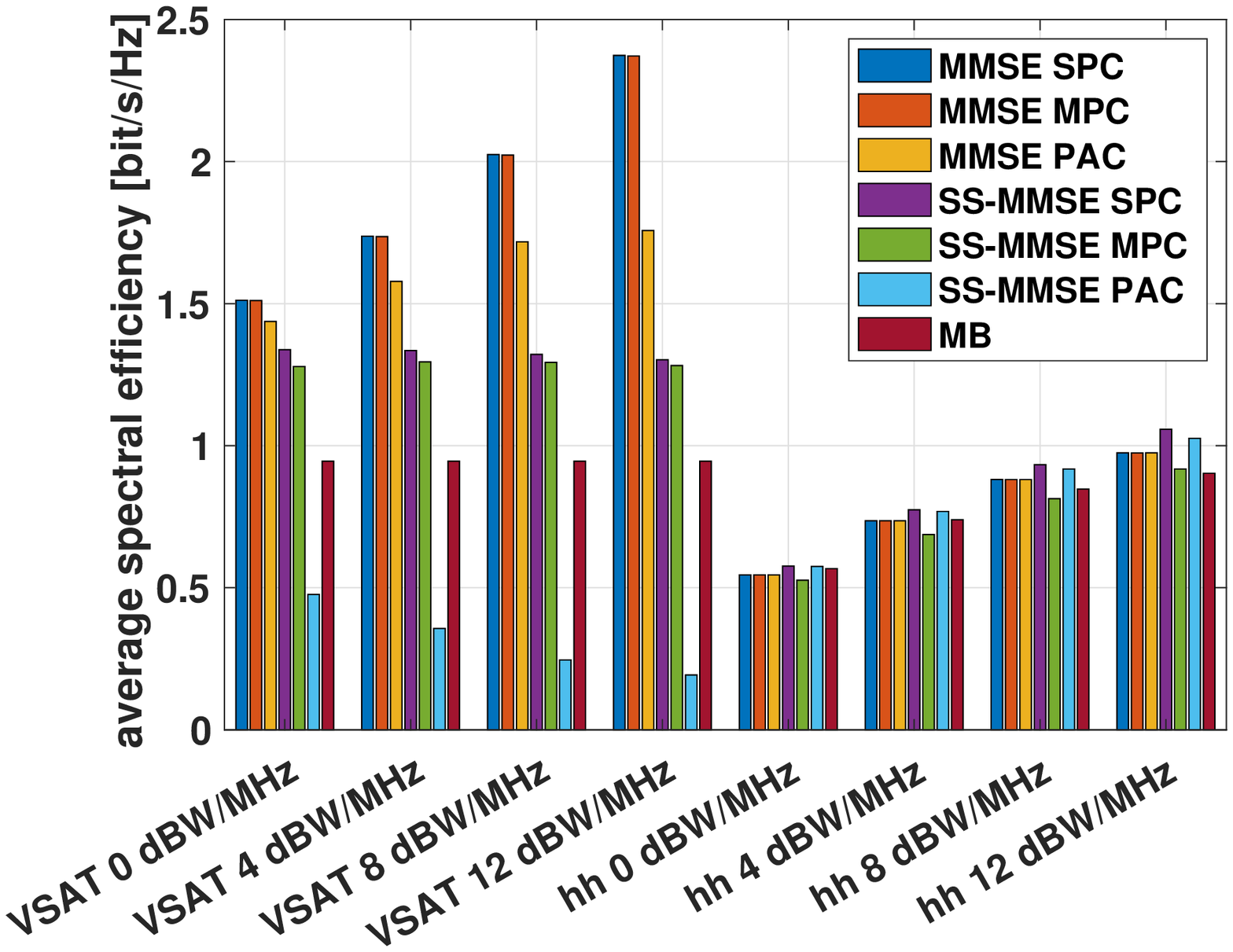}}
    \hfill
    \subfloat[Beam space \label{fig:nlosbeem}]{%
        \includegraphics[width=0.5\linewidth]{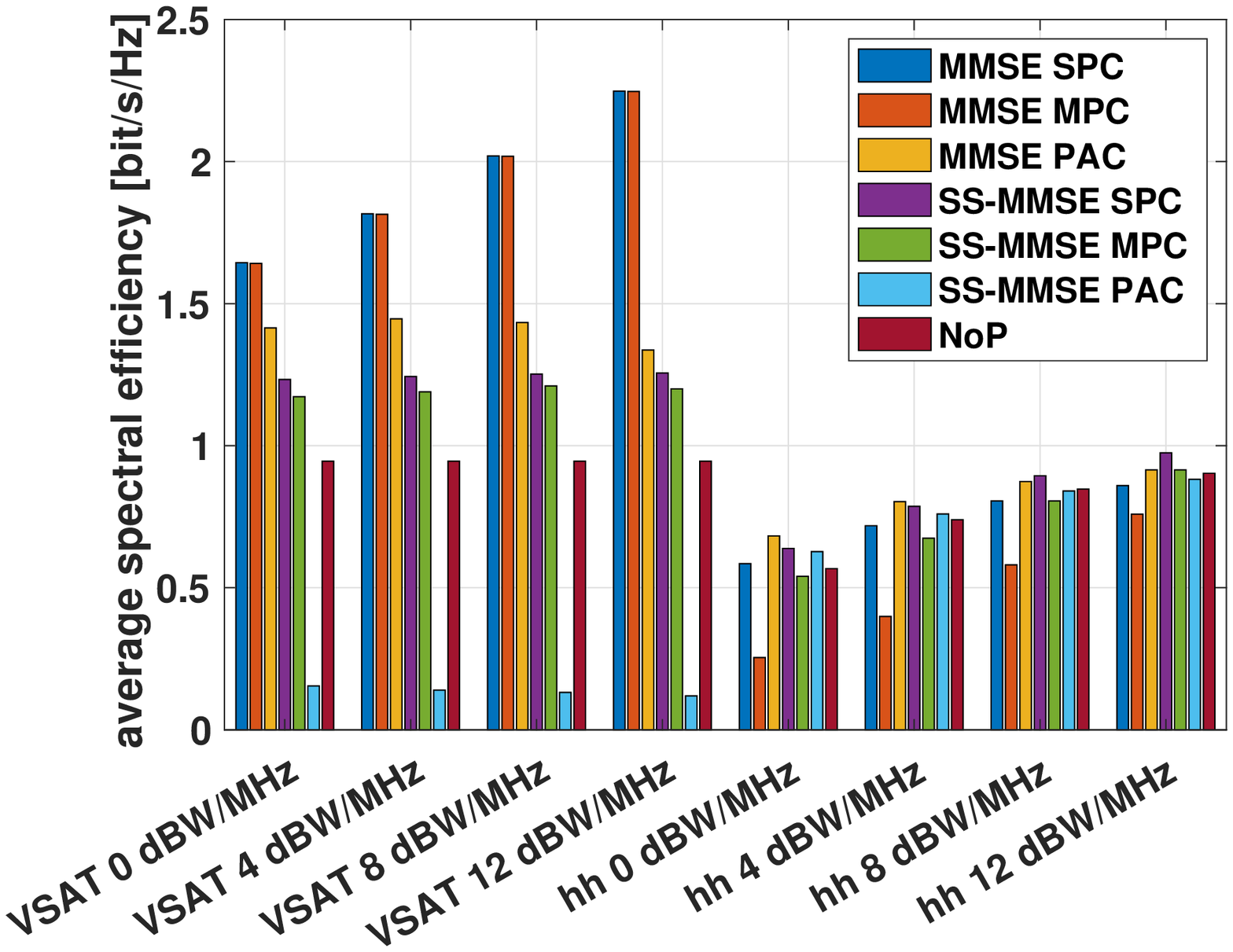}}
  \caption{Average Spectral efficiency of fixed users in NLoS scenario.}
  \label{fig:analysis} 
\end{figure}

To conclude the assessment for fixed terminals, we also consider NLOS propagation conditions in sub-urban environments. 
When the user is in NLOS conditions, in addition to the impairments already present for the pLOS scenario, it also experiences shadow fading, scintillation, gaseous absorptions, and Clutter Loss. 
Figures ~\ref{fig:nlosfeed} and ~\ref{fig:nlosbeem} provide the average spectral efficiency for the sub-urban environment in NLOS conditions, with feed and beam space precoding, respectively. In that case, the performance is significantly worse compared to beam and feed space precoding in pLOS conditions, with losses in the order of 2 bit/s/Hz and 4-5 bit/s/Hz, respectively. As already observed in the pLOS scenario, MMSE and SS-MMSE precoding with SPC and MPC normalisations improve the performance with larger power levels; while with the PAC normalisation, differently from the previous case, the MMSE precoding provides a good performance, relatively close to the MPC. Indeed, when including the clutter losses, the benefit of increasing the SNR is more impactful compared to the loss in the precoder orthogonality. This trend is not present for SS-MMSE precoding with PAC, which still shows a poor spectral efficiency; in this case, the further approximation of the channel matrix with that at beam center makes the SNR improvement negligible with respect to the orthogonality loss. With handheld terminals, the PAC approach is even better than the SPC. This behaviour is motivated by the extremely harsh propagation conditions which make the misalignment between the channel matrix and the precoding matrix significant. Consequently, with such large losses and without any gain at the receiver, it is better to equally allocate the power to the users, since the orthogonality is already disrupted. 

\subsubsection{Mobile terminals} 
\begin{figure}[t!]
    \centering
    \subfloat[Feed space \label{fig:plosfeed_safety}]{%
        \includegraphics[width=0.5\linewidth]{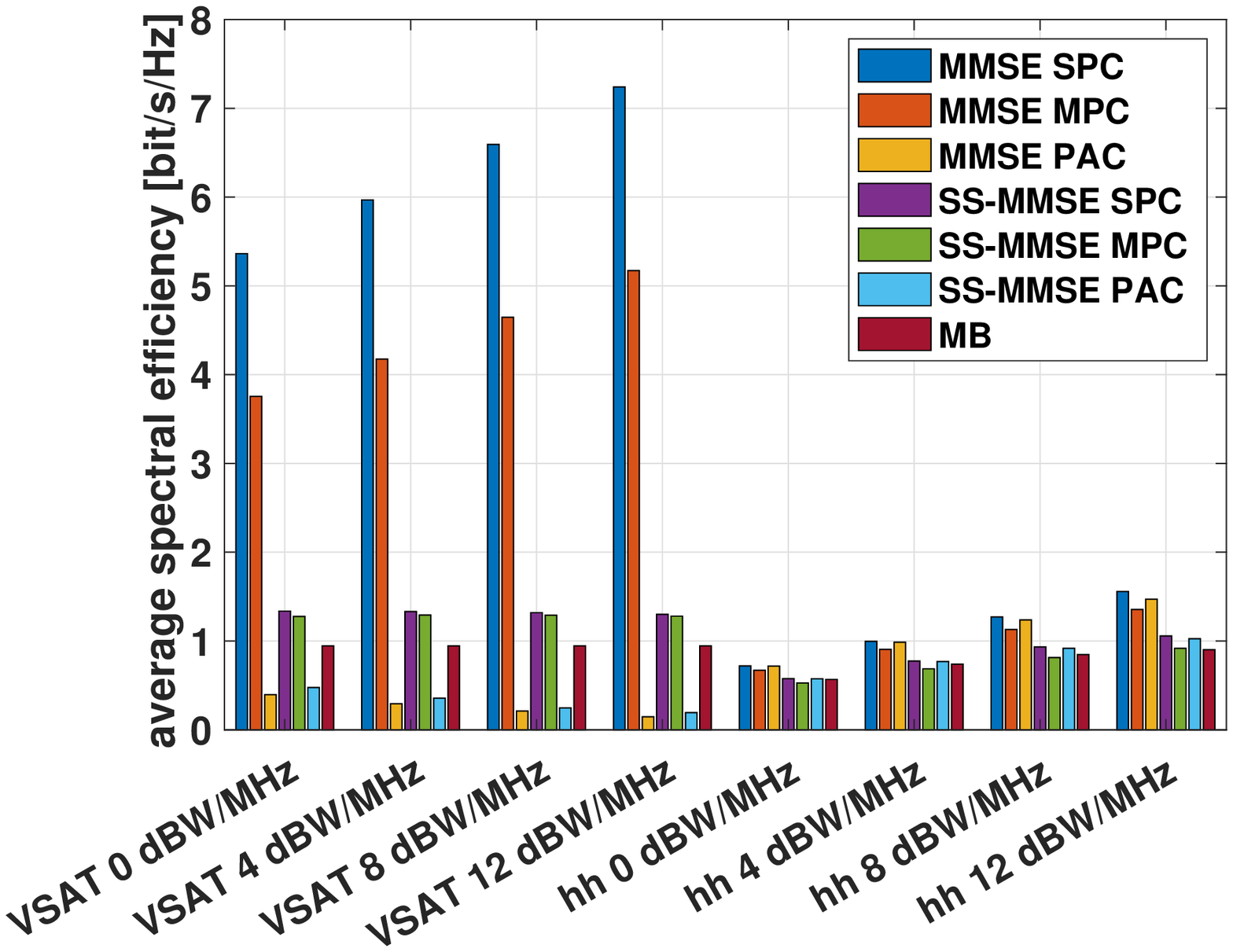}}
    \hfill
    \subfloat[Beam space\label{fig:plosfeed_beam}]{%
        \includegraphics[width=0.5\linewidth]{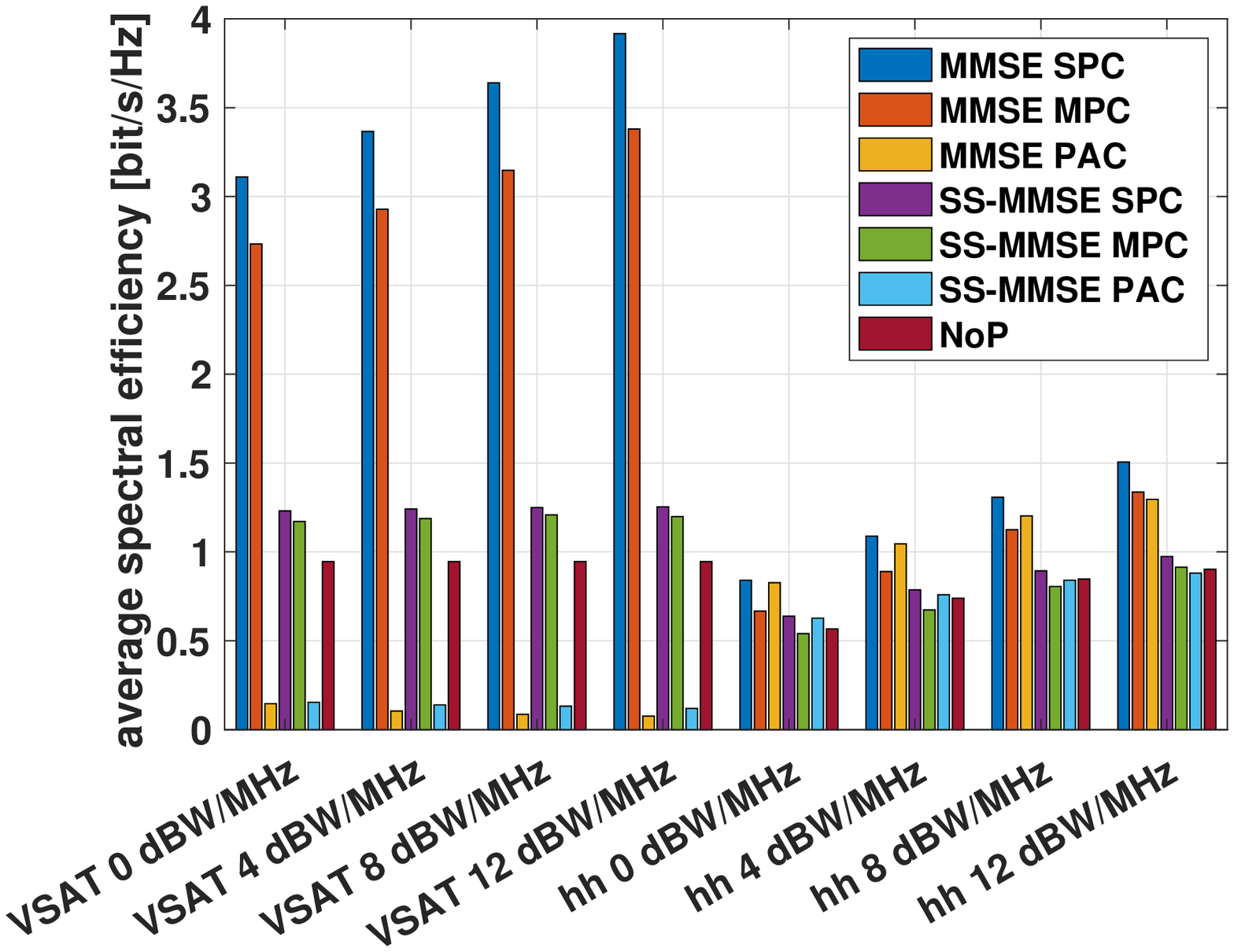}}
  \caption{Average Spectral efficiency of public safety terminals in pLoS scenario.}
  \label{fig:analysis} 
\end{figure}

\begin{figure}[t!]
    \centering
    \subfloat[Feed space \label{fig:nlosfeed_safety}]{%
        \includegraphics[width=0.5\linewidth]{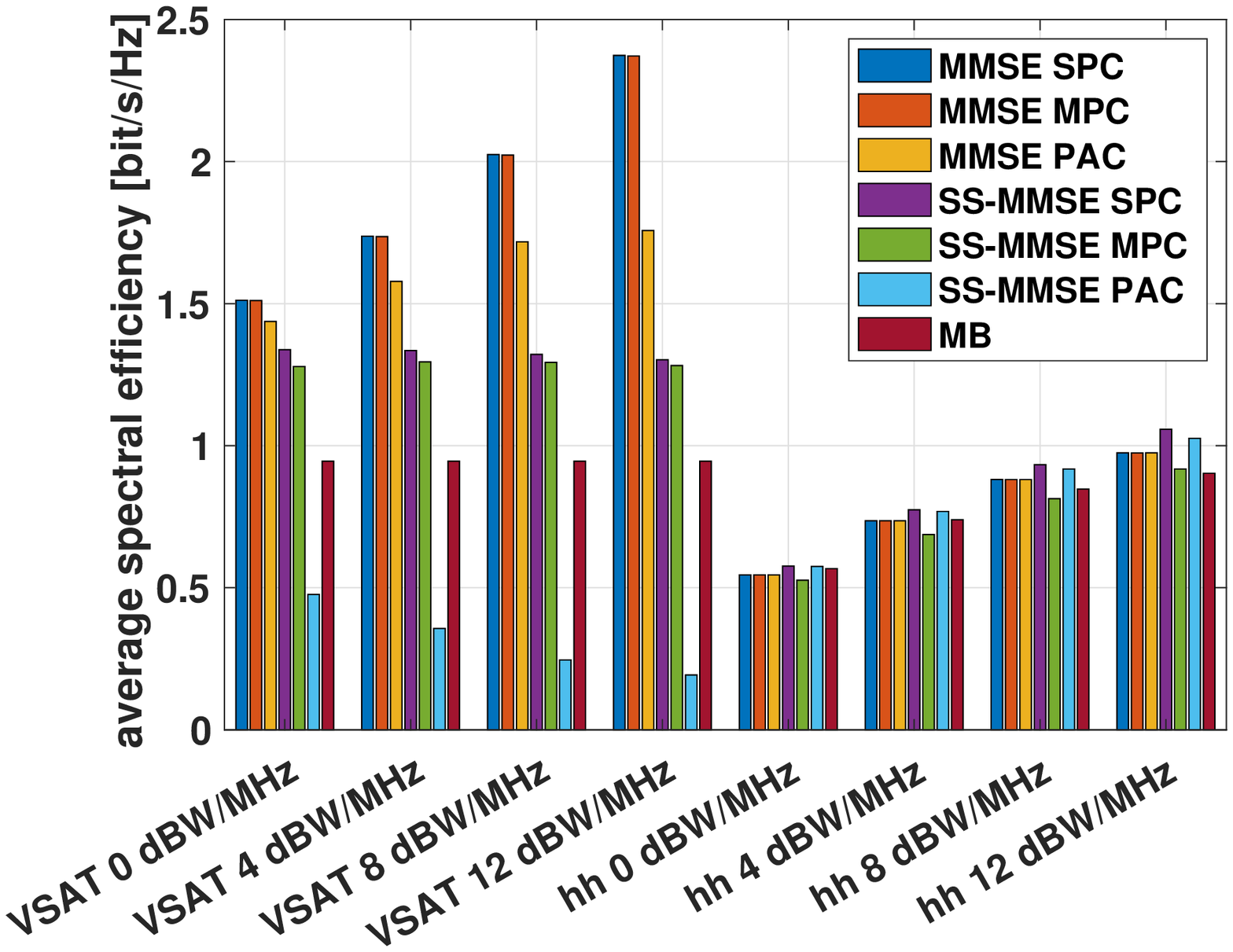}}
    \hfill
    \subfloat[Beam space\label{fig:nlosbeam_safety}]{%
        \includegraphics[width=0.5\linewidth]{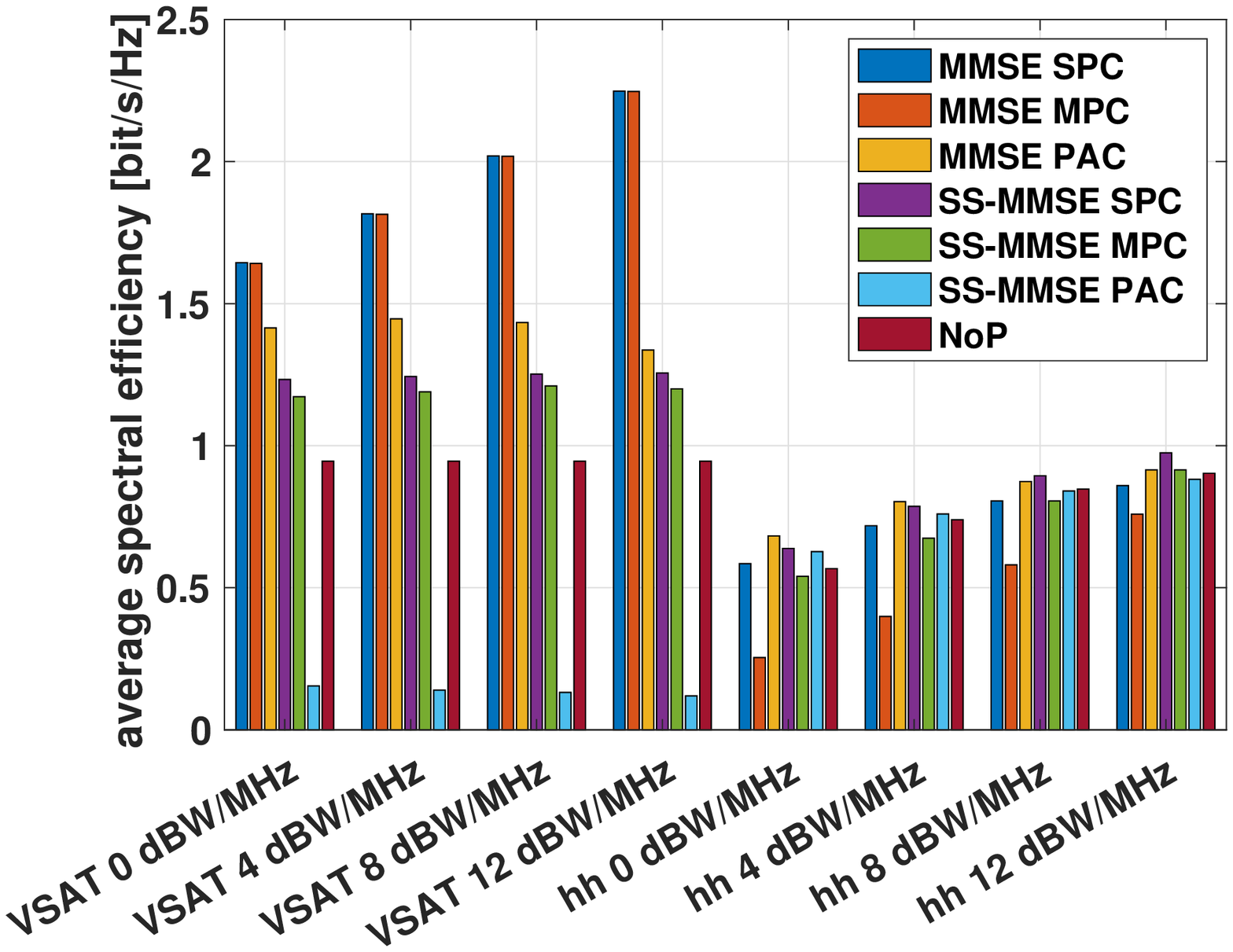}}
  \caption{Average Spectral efficiency of public safety terminals in NLoS scenario.}
  \label{fig:analysis} 
\end{figure}

in this scenario, public safety  terminals move at $v_{UE}=250 km/h$. In the limited time interval between the estimation and the transmission phase with CPC, which is expected to be even lower with a DPC architecture, where the precoding coefficients are computed on-board, there is a position error that leads to a further misalignment in the channel matrix used in the estimation phase and that in the transmission phase. It is also worth mentioning that this can be predicted by exploiting the known speed vector, with a small residual error. With this type of terminals, the distance travelled in this interval is equal to 1.156 meters. It is thus reasonable to expect that the impact of the users’ movement is negligible on the system performance compared to the other sources of non-ideal CSI (in particular the different realisations of the stochastic terms). 
For the sake of completeness, below we report the performance histograms in the beam and feed spaces for pLOS and NLOS propagation conditions in ~\ref{fig:plosfeed_beam},~\ref{fig:plosfeed_safety}, ~\ref{fig:nlosfeed_safety}, and ~\ref{fig:nlosbeam_safety}. By comparing these results with the corresponding histograms in the fixed terminal section, the Public Safety terminals provide a performance that is at most equal to that of fixed terminals or, in the worst case, with a spectral efficiency degradation in the order of $10^{-4} bit/s/Hz$, thus substantiating the above observations. 
\section{Conclusion}
In this work, we designed and assessed a precoding technique not requiring CSI at the transmitter, but based on location information (SS-MMSE), and compared it to CSI and non-CSI based benchmark algorithms (MMSE, MB).
Despite MMSE is always providing the best performance, SS-MMSE precoding shows an acceptable performance, also considering that it does not need a continuous reporting of CSI vectors. As for the normalisations, MPC and PAC provide, depending on the scenarios as discussed above, a performance close to that of SPC. They are to be preferred since they guarantee that each antenna feed is not emitting a transmission power above its maximum. Future works foresee the inclusion of distributed solutions with multiple satellites, tackling signalling aspects, and evaluating the performance at link level. 

\section{Acknowledgment}
This work has been funded by the European Union Horizon-2020 Project DYNASAT (Dynamic Spectrum Sharing and Bandwidth-Efficient Techniques for High-Throughput MIMO Satellite Systems) under Grant Agreement 101004145. The views expressed are those of the authors and do not necessarily represent the project. The Commission is not liable for any use that may be made of any of the information contained therein.

\bibliographystyle{IEEEtran}
\bibliography{IEEEabrv,biblio_RAML}

\end{document}